\documentstyle[12pt]{article}
\begin{document}
\setlength{\baselineskip}{14pt}

\title{Hyperfinite-operational Approach \\to the Problem of \\
Time Reversibility of Quantum Mechanics}
\author{Hideyasu Yamashita\thanks{Graduate School of Human Informatics, Nagoya
University, Nagoya 464-01, Japan. E-mail: test1@info.human.nagoya-u.ac.jp}}
\date{}

\maketitle
\sloppy

\begin{abstract}
This paper outlines a mathematical framework of quantum probability in 
which the time asymmetry in describing measuring processes is avoided. 
The main objects
of the framework are hyperfinite operations, which are constructed by using
nonstandard analysis and the operational approach by Davies and Lewis.
Then the notions of Bayesian conditional probability 
are defined,
and Bayes-type theorems in terms of the probability are showed.  
\end{abstract}

\def\N{{\bf N}}
\def\R{{\bf R}}
\def\C{{\bf C}}
\def\Z{{\bf Z}}
\def\M{{\bf M}}
\def\P{{\cal P}}

\section{Introduction}

Belinfante$^{(1)}$ argued the problem of retrodictions in quantum physics.
He insists: conventional quantum theory has only predictive concepts, e.g.,
predictive probability and state (he rather prefers the word ``postdictive,''
but its implication is similar to ``retrodictive'').
However, to complete the formalism of quantum mechanics we need a satisfactory
treatment of retrodictive concepts.

This proposal leads to his program of ``time-symmetric quantum theory'' 
including quantum measurement theory. This program seems to be of great 
interest in two respects; one is theoretical, and the other is philosophical.

The first point is coherence of the formalism of quantum measurement theory
(usually nonrelativistic) and that of relativistic quantum theory$^{(1,2)}$.
 Roughly speaking, the way
in which quantum measurement theory treats the time coordinate and the way
in which it treats the space coordinates are radically different. A great
part of the difference is caused by the time asymmetry of quantum measurement
theory.

The second point is the problem of interpretation of state reduction, e.g.,
Schr\"{o}dinger's cat paradox. The time asymmetry appears mainly in
descriptions of state reduction\footnote{Sometimes ``state reduction'' is
used to refer a transformation from a pure state to a mixed state$^{(4)}$,
sometimes to refer state transformation $|\psi\rangle\rightarrow P|\psi
\rangle$ where $|\psi\rangle$ is a vector in Hilbert space ${\cal H}$
and $P$ is a projector on ${\cal H}$. Both of them imply time asymmetry
though those usage must be distinguished.}. 
Ozawa$^{(5)}$ gave a new formulation to the problem in the framework of $C^*$-
dynamical systems. He proved that in the framework we cannot describe the
measuring interaction between a microscopic system and a macroscopic
apparatus as long as the time reversibility of the dynamics of isolated 
system is assumed. 

The present paper proposes a mathematical framework of time-symmetric quantum
physics. This framework uses two tools: the operational approach to quantum
probability by Davies and Lewis$^{(6,7)}$, and the method of nonstandard
analysis originated by A.Robinson$^{(8)}$. Our idea of the resolution
of time asymmetry is similar to that of Belinfante's; 
conventional quantum measurement theory has only predictive concepts, 
so it has time asymmetry. Therefore, we should add retrodictive concepts to
it. We shall introduce two types of Bayesian conditional probability
(predictive and retrodictive) in section 4. These two types  
of probability will be related by Bayes-type theorems.

\section{Nonstandard Analysis}

This section briefly outlines the theory of nonstandard analysis$^{(5)}$.
Let $X$ be a set and ${\cal P}(X)$ the power set of $X$, that is, the set 
of all subsets of $X$. The {\it superstructure over} $X$, denoted by $V(X)$,
 is 
defined by the following recursion:
\[V_0(X)=X, V_{n+1}(X)=V_n(X)\cup{\cal P}(V_n(X)),\]
\[V(X)=\bigcup _{n\in {\bf N}}V_n(X),\]
where ${\bf N}$ is the set of natural numbers. Let us regard any element of 
$X$ as a nonset here; hence $x\in V(X)$ is a set iff $x\in V(X)\setminus X$. 
Let 
${\bf C}$ be the set of complex numbers. $V({\bf C})$ contains all the 
structures that we use in quantum physics; for instance, separable Hilbert 
space 
${\cal H}$ is in $V({\bf C})$.

$V(X)$ is called a {\it nonstandard extension of\/} $V({\bf C})$
 if there exists a map $\star :V({\bf C})\longrightarrow V(X)$ satisfying 
the following conditions:
\begin{list}{}{}
\item (1)  $\star $ is an injective mapping from $V({\bf C})$ to $V(X)$,
\item (2)  $^\star {\bf C}=X$,
\item (3) (Transfer Principle) Let $\phi$ be a sentence in terms of 
$V({\bf C})$, and $^\star\phi $ be the sentence ``transfered'' from $\phi$ 
by mapping $\star.$ $\phi$ is true iff $^\star\phi$ is true.
\end{list}

\begin{sloppypar}
Transfer Principle needs more explanation. A sentence in terms of $V({\bf C})$
 is constructed from the symbols for logical connectives $\neg, \wedge, \vee, 
\Rightarrow, \Leftrightarrow$, quantifiers $\forall, \exists$, individual 
variables $x_1, x_2,...$ , two predicates $=,\in$, parentheses $(, )$, and 
elements of $V({\bf C})$.

We will consider an example. Let {\bf R} denote the set of real 
numbers. Define $G_<\in V({\bf C})$ by 
$G_<=\{(x,y)|\,x,y\in {\bf R}, x<y\}$, where $(x,y)$ 
is identified as $\{\{x\},y\}$. 
\end{sloppypar}
\[(\forall x)(\forall y)(x\in{\bf R}\wedge y\in{\bf R}\wedge (x,y)\in G_<
\Rightarrow(\exists z)(z\in{\bf R}\]
\[\wedge (x,z)\in G_<\wedge (x,y)\in G_<))\]
 is a sentence in terms of $V({\bf C})$ because ${\bf R},G_<\in V({\bf C})$.
 Let $\phi$ denote this sentence. $\phi$ means that ${\bf R}$ is dense, 
and hence $\phi $ is true. The ``transfered'' sentence $^\star\phi$ is as 
follows:
\[(\forall x)(\forall y)(x\in\!^\star{\bf R}\,\wedge \,y\in\!^\star{\bf R}
\,\wedge\,(x,y)\in\! ^\star G_<\Rightarrow(\exists z)(z\in\!^\star{\bf R}\]
\[\wedge\, (x,z)\in\! ^\star G_<\,\wedge \,(x,y)\in\! ^\star G_<))\]
By Transfer Principle, $^\star\phi$ is true  ($^\star{\bf R}$ is called 
$\star$-dense).

$u\in V(X)$ is called {\it standard\/} if there is $x\in V({\bf C})$ such that
 $u=$$^\star x$, and called {\it internal\/} if there is $x\in V({\bf C})$ 
such 
that $u\in$$^\star x$. $^\star V({\bf C})$ is the set of all internal sets. 
Let $A$ and $B$ be internal sets.
Function $f:A\longrightarrow B$ is called {\it internal\/} if the graph of $f$,
that is, $\{(x,f(x))|x\in A\}$ is internal. $V(X)$ is called a {\it countably 
saturated extension of\/} 
$V({\bf C})$ if it satisfies the following condition:

(Saturation Principle) If countable sequence of internal sets $A_j\in V(X)
\setminus X$ satisfies
\[\bigcap_{j=1}^kA_j\neq\phi\,(k=1,2,...)\]
then
\[\bigcap_{j=1}^{\infty}A_j\neq\phi.\]

Let $A\in V({\bf C})\setminus {\bf C}$. From the Saturation Principle, we can 
show that if $A\in V({\bf C})$ is an infinite set, $^sA$ defined by 
\[^sA=\{^\star a|\,a\in A\}\]
is a proper subset of $^\star A$. Thus $^\star A$ is an extended structure
 of $A$. 
$^\star A$ is called the {\it nonstandard extension of\/} A. 
By renaming the elements of $X$,
 we can assume without loss of generality that ${\bf C}$ is a subset of $^\star
 {\bf C}$ and $^\star x=x$ for each $x\in {\bf C}$.

Given any subset $U\subseteq V(\C)$, define $^\star U$ by
\[^\star U=\bigcup_{n\in {\bf N}}\!^\star(U\cap V_n(\C)).\]
Let $F({\bf C})\subseteq V({\bf C})$ be the set of finite sets, i.e.,
\begin{center}
$F({\bf C})=\{A\in V({\bf C})\setminus{\bf C}|\,A$ is a finite set\}.
\end{center}
An element of $^\star F({\bf C})$ is called a {\it hyperfinite set\/}. 
Any element of $^\star {\bf C}\,(^\star {\bf R})$ is called a {\it 
hypercomplex (hyperreal)\/} number. We assume that $^\star 
{\bf R}\subseteq\!^\star {\bf C}$.
 $^\star{\bf C}$ is a proper extension field of ${\bf C}$, 
 and $^\star {\bf R}$ is an ordered extension of ${\bf R}$.
An element of $^\star {\bf N}\,(^\star {\bf Z})$ is called a {\it 
hypernatural number (hyperinteger)\/}. It is shown that if $A$ is a
hyperfinite set (i.e. $A\in\!^\star F({\bf C})$) then there
is an initial segment $J=\{n\in\!^\star{\bf N}|\,n\le j\}$ for some $j\in\!
^\star {\bf N}$ and a one-to-one, onto internal mapping $f:J\longrightarrow A$.
Thus, we will often write a hyperfinite set $A$ as $A=\{a_1,a_2,...,a_j\}
 $, where $a_k=f(k),k\in J$.

A hypercomplex number $x$ is called {\it infinite\/} if $|x|>n$ for any $n\in 
{\bf N}$, {\it finite\/}, $|x|<\infty$, if there is some $n\in {\bf N}$ such
 that $|x|<n$, and {\it infinitesimal\/} if $|x|<\frac{1}{n}$ for any $n\in
{\bf N}$.

For any $x,y\in \!^\star {\bf C}$, we will write $x\approx y$ if $|x-y|$ is
infinitesimal. For any finite hyperreal number $x$, there is a unique real
number $r$ such that $^\star r\approx x$; this $r$ is called the {\it standard
part of\/} of $x$ and denoted by $^\circ x$.
 
Any function $f$ from $A$ to $B$ is extended to an internal function $^\star f$
 from $^\star A$ to $^\star B$. A sequence $a_n\in{\bf C}\,(n\in \N)$
 is extended to an internal sequence $^\star a_\nu\in\!^\star{\bf C}\,(\nu
\in\!^\star {\bf N})$, so that $\lim _{n\to\infty}a_n=a$ if and only if
$^\star a_\nu \approx a$ for all $\nu \in \!^\star{\bf N}\setminus {\bf N}$.

\def\fin{\mathop{\rm fin}\nolimits}
Let $\cal A$ be an internal normed linear space with norm $\parallel \cdot
\parallel$. The {\it principal galaxy\/} $\fin({\cal A})$ and the {\it 
principal monad\/} $\mu(0)$ are defined by
\[\fin({\cal A})=\{x\in{\cal A}|\,\parallel x\parallel<\infty\}\]
\[\mu(0)=\{x\in{\cal A}|\,\parallel x\parallel \approx 0\}\]
 Both of them are linear spaces over $\C$. The {\it nonstandard hull of\/}
${\cal A}$ is the quotient linear space $\hat{{\cal A}}=\fin({\cal A})/\mu(0)$
equipped with the norm given by
\[\parallel ^\circ x\parallel=\!^\circ \parallel x\parallel\]
for all $x\in \fin({\cal A})$, where $^\circ x=x+\mu(0)$. It is shown
by the Saturation Principle that $\hat{\cal A}$ is a Banach space(Ref.(9),
 p.155).

Let $\nu$ be an infinite hypernatural number, i.e., $\nu\in\! ^\star \N
\setminus
\N$, and $^\star\C^\nu$ the $\nu$-dimensional internal inner product space
with the natural inner product and the internal norm $\parallel\cdot\parallel
$ derived by the inner product. Let $\M=\!^\star M(\nu)$ be the internal
algebra of $\nu\times\nu$ matrices over $^\star\C$. Naturally, $\M$ acts
on $^\star\C^\nu$ as the internal linear operators, and let $p_{\infty}$
be the operator norm on $\M$, i.e., $p_{\infty}(A)=\sup\{\parallel A\xi
\parallel |\,\parallel\xi\parallel\le 1, \xi\in\!^\star\C^\nu\}$. Denote
by $A^*$ the adjoint of $A\in\M$. Let $\tau$ be the internal normalized
trace on $\M$, i.e., 
\[\tau (A)=\frac{1}{\nu}\sum_{i=1}^{\nu}A_{ii}\]
for $A=(A_{ij})\in\M$. Then $\tau$ defines an internal inner product
$(\cdot|\cdot)$ on $\M$ by $(A|B)=\tau(A^*B)$, for $A,B\in\M$. Its derived
norm called the normalized Hilbert-Schmidt norm is denoted by $p_2$, i.e.,
$p_2(A)=\tau(A^*A)^{1/2}$ for $A\in \M$. 
Denote by $(\M,p_\infty)$ and $(\M,p_2)$ the normed linear spaces equipped
with these respective norms. The principal galaxies $\fin_{\infty}(\M)$ of
$(\M,p_{\infty})$ and $\fin_2(\M)$ of $(\M,p_2)$ are defined as follows:
\[\fin_{\infty}(\M)=\{A\in\M|\,p_{\infty}(A)<\infty\}\]
\[\fin_{2}(\M)=\{A\in\M|\,p_{2}(A)<\infty\}\]
The principal monads $\mu_{\infty}(0)$ of $(\M,p_{\infty})$ and $\mu_2(0)
$ of $(\M,p_2)$ are defined as follows:
\[\mu_{\infty}(0)=\{A\in\M|\,p_{\infty}(A)\approx  0\}\]
\[\mu_{2}(0)=\{A\in\M|\,p_{2}(A)\approx  0\}.\]
The nonstandard hull $\hat\M_2=\fin_2(\M)/\mu_2(0)$ turns out to be a Hilbert
space with inner product $<\cdot|\cdot>$ and norm $\parallel\cdot\parallel
_2$ defined by 
\[<A+\mu_2(0)|B+\mu_2(0)>=\!^{\circ}(A|B)\]
and
\[\parallel A+\mu_2(0)\parallel_2=\!p_2(A).\]
for $A,B\in\fin_2(\M)$.

The nonstandard hull $\hat{\M}_{\infty}=\fin_{\infty}(\M)/\mu_{\infty}(0)$
of $(\M,p_{\infty})$ turns out to be a $C^*$-algebra equipped with norm
$\hat p_{\infty}$ defined by 
\[\hat p_{\infty}(A+\mu_{\infty}(0))=\!^\circ p_{\infty}(A)\]
 for $A\in\fin_{\infty}(\M)$.

Hinokuma and Ozawa$^{(10)}$ showed that another quotient space $\hat\M$ defined
by
\[\hat\M=\fin_{\infty}(\M)/(\mu_2(0)\cap\fin_{\infty}(\M)).\]
is a von Neumann algebra of type I\rm{I}$_1$ factor.

\def\tr{\mathop{\rm tr}\nolimits}

\section{Operation}

This section briefly reviews the notion of operation and instrument in the
sense of Davies ,Lewis, Srinivas and Ozawa$^{(6,7,11,12)}$.

The state space $V$ of a Hilbert space ${\cal H}$ is defined as the Banach
space ${\cal T}_s({\cal H})$ of trace class operators on ${\cal H}$ with trace
norm $\parallel\cdot\parallel _{\tr}\,(=\tr|\cdot|)$. The states are defined as
the non-negative trace class operators $V^+$ of trace one, elsewhere called
density matrix.

If $V$ is the state space of Hilbert space ${\cal H}$, an {\it operation\/}
on $V$ is defined as a positive linear map $T:V\rightarrow V$ which also
satisfies
\[0\le\tr(T\rho)\le\tr \rho.\]
for all $\rho\in V^+.$ The concept of operation is thought of as an extension
of that of projective observations. Let $P$ be a projector on ${\cal H}$
and $\rho$ a state. The probability that the ``proposition'' $P$ is observed
by a projective observation is $\tr(P\rho)$ and the state at the instant 
after the observation is $P\rho P/\tr(P\rho )$. If $T$ denote the map
$T:\rho\mapsto P\rho P$, then the probability  is written as $\tr(T\rho)$
and the state as $T\rho/\tr(T\rho).$ $T$ is an example of an operation.

An {\it instrument\/} is defined as a mapping
\[I:B(R)\times V\rightarrow V\]
where $B(R)$ is the set of all Borel sets of a value space $R$ (usually
a real line). The requirements on $I$ for it to define an instrument are
\begin{list}{}{}
\item (i) $I(E,\cdot)$ is an operation for all $E\in B(R)$.
\item (ii)$I(\bigcup_i E_i,\rho)=\sum_i I(E_i,\rho)$ for each countable
family $\{E_i\}$ of pairwise disjoint sets of $E_i\in B(R)$.
\item (iii)$\tr I(R,\rho)=\tr\rho.$
\end{list}
 Davies and Lewis proved that given two instruments $I_1$ and $I_2$ on
value spaces $R_1$ and $R_2$, we can consider their composition $I_2(E_2,I_1(E_1,\rho))$
as a unique instrument $I$ defined on $R_1\times R_2$.

Ozawa$^{(12)}$ introduced the concept of realizability of an operation, 
and showed
that operation $T$ is realizable iff $T$ is completely positive, that is, 
for any finite sequence of vectors $\xi_1,...,\xi_n$ and $\eta_1,...,
\eta_n,$
\[\sum_{i,j=1}^n\langle\xi_i|T(|\eta_i\rangle\langle\eta_j|)|\xi_j\rangle
\ge 0.\]
Any operation which is not realizable is of no interest from a physical
point of view, so we may add complete positivity to the definition of operation
.

\section{Hyperfinite Operation}
\def\LM{\!^\star{\cal L}({\bf M})}

Let $\LM$ denote the set of all internal linear mapping $a:\M\rightarrow\M.$

{\bf Definition 4.1} Let $a\in\LM$. $a^{\leftrightarrow},a^{\updownarrow},
a^{\diamond}\in\LM$ are defined as follows:

\[a^{\leftrightarrow}(A)=[a(A^*)]^*,\]
\[\tr[B^*a^{\updownarrow}(A)]=\tr[A^*a(B)] \]
for all $A,B\in\M$, and
\[\langle\gamma|a^{\diamond}(|\alpha\rangle\langle\beta|)|\delta\rangle=
\langle\gamma|a(|\delta\rangle\langle\beta|)|\alpha\rangle\]
for any pairwise orthogonal vectors $|\alpha\rangle,|\beta\rangle,|\gamma
\rangle,$ and $|\delta\rangle$.

\def\us{\!{}^{\star}}
\def\ud{{}^{\updownarrow}}
\def\lr{{}^{\leftrightarrow}}
\def\dia{{}^{\diamond}}

Suppose $A\in\M$ and $a\in\LM$. If we represent $A$ as $(A_{\alpha\beta})$, $
a(A)$ is represented as 
\[\Bigl(\sum_{\alpha\beta}a_{\gamma\delta}^{\alpha\beta}A_{\alpha\beta}
\Bigr)_{\gamma\delta},\]
and hence $a$ is denoted as
\[(a_{\gamma\delta}^{\alpha
\beta}).\]
 Under this denotation, we see that
\[(a\lr)_{\gamma\delta}^{\alpha\beta}=\overline{a_{\delta\gamma}^{\beta\alpha}}
\] 
\[(a\ud)^{\alpha\beta}_{\gamma\delta}=
\overline{a_{\alpha\beta}^{\gamma\delta}}\]
\[(a\dia)_{\gamma\delta}^{\alpha\beta}=a_{\gamma\alpha}^{\delta\beta}\]
For $a,b\in\LM,$ let $ab$ denote the composed mapping of $b$ and $a$. Thus,
 the following properties are clear:
\begin{list}{}{}
\item (i)  $a\lr\lr =a\ud\ud =a\dia\dia =a,$
\item (ii) $a\lr\dia =a\dia\ud,$ $a\dia\lr=a\ud\dia,$
\item (iii)$(ab)\ud =b\ud a\ud, (ab)\lr =a\lr b\lr,$

\item (iv) $a\ud\lr =a\lr\ud$
\end{list}
 
{\bf Definition 4.2} Suppose $a\in\LM$. $a$ is called {\it positive\/} if 
for some
$b\in\LM,a=b\ud b$. 

This definition is an analogy of that of positive matrix; a matrix
$A$ is called {\it positive\/} if $A=B^*B$ for some matrix $B$.
 
{\bf Proposition 4.1}  Suppose $a\in\LM$. Following three conditions are
equivalent.
\begin{list}{}{}
\item (i)   $a$ is positive.
\item (ii)  $\tr[A^*a(A)]\ge0$ for each $A\in \M$.
\item (iii) $\langle\alpha|a(|\alpha\rangle\langle\beta|)|\beta\rangle\ge 0$ 
for all vectors $|\alpha\rangle,|\beta\rangle\in\!^\star\C^{\nu}$.
\end{list}

{\bf Proof}  Evident from the fact that $<A,B>=\tr(A^*B)$ is an inner product
on $\M$,and $a\ud$ is the adjoint operator of $a$ on the inner product space
$(\M,<,>).\hspace{5mm}\Box$

{\bf Proposition 4.2}  Suppose $a\in\LM$. Following four conditions are 
equivalent.
\begin{list}{}{}
\item (i)   $a\dia$ is positive.
\item (ii)  $\tr[Aa(B)]\ge0$ for all positive $A,B\in\M$.
\item (iii) $a(A)$ is a positive matrix for any positive $A\in\M$.
\item (iv)  $\langle\alpha|a(|\beta\rangle\langle\beta|)|\alpha\rangle$
$\ge 0$ for all $|\alpha\rangle,|\beta\rangle\in\!^\star\C^{\nu}$.
\end{list}

We see from (iii) that if $a\dia$ and $b\dia$ are positive, $(ab)\dia$
is positive.

{\bf Proposition 4.3}   Suppese $a\in\LM.$ $a\dia$ is positive iff there
are $\kappa\in\us\N$ and $M_k\in\M\,(k=1,\cdots,\kappa)$ such that
\[a(A)=\sum_{k=1}^{\kappa}M_kAM_k^*.\]

{\bf Proof}   If $a(A)=\sum_{k=1}^{\kappa}M_kAM_k^*$, $a\dia$ is positive
by Proposition 4.2 (iii). Conversely, if $a\dia$ is positive, there is
$b\in\LM$ such that $a\dia =b\ud b$, and hence we have 
\[a_{\gamma\delta}^{\alpha\beta}=\sum_{\epsilon,\zeta}\overline{(b\dia)
_{\epsilon\gamma}^{\zeta\alpha}}(b\dia)_{\epsilon\delta}^{\zeta\beta}.\]
If $B^{\alpha}_{\beta}\in\M\,(\alpha,\beta=1,\cdots,\kappa)$ are defined by
\[(B_{\beta}^{\alpha})_{ij}=(b\dia)_{\beta j}^{\alpha i},\]
then,
\begin{eqnarray*}
&&[a(A)]_{\gamma,\delta}=\sum_{\alpha,\beta}a_{\gamma\delta}^{\alpha\beta}
A_{\alpha\beta}=\sum_{\alpha,\beta,\epsilon,\zeta}\overline{(b\dia)_{\epsilon
\gamma}^{\zeta\alpha}}A_{\alpha\beta}(b\dia)_{\epsilon\delta}^{\zeta\beta}\\
&&=\Bigl(\sum_{\zeta,\epsilon}B_{\epsilon}^{\zeta *}AB_{\epsilon}
^{\zeta}\Bigr)_{\gamma,\delta}.
\end{eqnarray*}
Thus $a(A)=\sum_{\zeta,\epsilon}B_{\epsilon}^{\zeta *}AB_{\epsilon}^{\zeta}
.\hspace{5mm}\Box$

{\bf Corollary\/} Let $a\in \LM$. $a\dia$ is positive iff $a$ is $\star$-
completely positive, i.e., for any hyperfinite sequence of vectors 
$|\xi_1\rangle,...,|\xi_{\kappa}\rangle,|\eta_1\rangle,...,|\eta_{\kappa}
\rangle\in\us\C^{\nu}$,
\[\sum_{i,j=1}^{\kappa}\langle\xi_i|T(|\eta_i\rangle\langle\eta_j|)|\xi_j
\rangle\ge 0\]
 
{\bf Definition 4.3}   Let $a\in\LM$. Linear functionals $\tr\ud,\tr\lr$ over
$\LM$ are defined as follows:
\[\tr\ud a=\sum_{\alpha,\beta}a_{\alpha\beta}^{\alpha\beta},\]
\[\tr\lr a=\sum_{\alpha,\beta}a_{\beta\beta}^{\alpha\alpha}.\]
 Evidently, we have $\tr\ud(ab)=\tr\ud(ba)$, while $\tr\lr(ab)=\tr\lr (ba)$
does not hold.

{\bf Proposition 4.4}   Let $I\in\M$ denote the identity matrix.
\begin{list}{}{}
\item (i)   $\tr\lr(a\dia)=\tr\ud a.$
\item (ii)  $\tr\ud(a\dia)=\tr\lr a.$
\item (iii) $\tr\lr a=\tr a(I).$
\item (iv)  $\tr\ud a=\tr a\dia(I).$
\item (v)   $\tr\lr(ab)=\tr[a\ud (I)^*b(I)]$
\end{list}

{\bf Proof}   (v). Let $\delta_{\alpha\beta}$ be Kronecker's notation.
\[\tr\lr (ab)=\sum_{\alpha,\beta,\gamma,\delta,\epsilon,\zeta}\delta
_{\alpha\beta}a_{\alpha\beta}^{\gamma\delta}b_{\gamma\delta}^{\epsilon\zeta}
\delta_{\epsilon\zeta}=\sum_{\gamma,\delta}\overline{(a\ud(I))_{\gamma,\delta}}
(b(I))_{\gamma,\delta}\]
\[=\tr[a\ud(I)^*b(I)].\hspace{5mm}\Box\]

{\bf Proposition 4.5}   Let $a\in\LM$. Suppose $a\dia$ be positive. If 
\[a(A)=\sum_{k=1}^{\kappa}M_kAM_k^*\,(\kappa\in\us\N,M_k\in\M)\]
 for all $A\in\M,$
 then,
 \[a\ud(A)=\sum_{k=1}^{\kappa}M_k^*AM_k\]
 for all $A\in\M$.

{\bf Proposition 4.6} Let $a,b\in\LM$.
\begin{list}{}{}
\item (i)   If $a\dia$ is positive, $\tr\lr a\ge 0$
\item (ii)  If $a\dia$ and $b\dia$ are positive and $a(I)\le I$, then
 $\tr\lr(ab)\le\tr\lr b$.
\item (iii) If $a\dia$ and $b\dia$ are positive, and $a\ud(I)\le I$, then
$\tr\lr(ba)\le\tr\lr b$.
\end{list}

{\bf Proof} (i). By Proposition 4.2(iii), $a(I)$ is positive matrix. Hence,
by Proposition 4.4(iii), $\tr\lr a=\tr[a(I)]\ge 0$. (ii). By Proposition 
4.2(iii)
 and 4.4(v), $\tr\lr(ab)=\tr[a\ud(I)b(I)]\le\tr  b(I)=\tr\lr b$. (iii) is 
shown in a sinilar way.$\hspace{5mm}\Box$

{\bf Definition 4.4} $a\in\LM$ is called a {\it hyperfinite operation\/}, or
simply an {\it operation\/} if $a\dia$ is positive
and $a(I),a\ud(I)\le I$. 

Let $Op$ denote the set of all operations.
$Op$ is an internal semigroup with involution $\updownarrow$ because if
$a,b\in Op,$ then $a\ud\dia$ and $(ab)\dia$ are positive, and 
$a(b(I))\le a(I)\le I$.

{\bf Proposition 4.7} $a\in\LM$ is an operation iff there exist $M_1,\cdots
,M_{\kappa}\in\M\,(\kappa\in\us\N)$ such that for all $A\in\M$,
\[a(A)=\sum_{k=1}^{\kappa}M_kAM_k^*\]
and
\[\sum_{k=1}^{\kappa}M_k^*M_k\le I,\,\,\sum_{k=1}^{\kappa}M_kM_k^*\le I.\]

{\bf Proof} Suppose $a\dia$ is positive. By Proposition 4.3 and 4.5, there
are $M_1,...,M_{\kappa}\in\M\,(\kappa\in\us\N)$ such that
\[a(A)=\sum_{k=1}^{\kappa}M_kAM_k^*,\]
\[a\ud(A)=\sum_{k=1}^{\kappa}M_k^*AM_k.\]
Hence, $a(I)\le I$ iff $\sum_{k=1}^{\kappa}M_kM_k^*\le I$, and $a\ud(I)\le I$ 
iff $\sum_{k=1}^{\kappa}M_k^*M_k\le I.\hspace{5mm}\Box$

{\bf Definition 4.5} If $P\in\M$ is a projector, i.e., $P^*=P^2=P$,
 $p:A\mapsto PAP$ is called a {\it projecting operation\/}. If $U\in\M$ is 
a unitary matrix, $u:A\mapsto
UAU^*$ is called a {\it unitary operation}, and $u^{-1}$ is defined as
$u^{-1}:A\mapsto U^*AU$. The identity $1\in Op$ is called a {\it unit operation
}. The zero element of $Op$ is written as $0$, i.e., $0:A\mapsto O$ for any
matrix $A$. Operation $a$ is called {\it trivial\/} if for any $b\in Op$,
$\tr\lr(ab)=\tr\lr(ba)=\tr\lr b$.

\def\rP{\stackrel{\rightarrow}{P}}
\def\lP{\stackrel{\leftarrow}{P}}

{\bf Definition 4.6} Let $a,b$ be operations and $b\neq 0$ (and hence
$\tr\lr b>0$).
\[\lP (a|b)=\!^\circ\left(\frac{\tr\lr(ab)}{\tr\lr b}\right)\]
\[\rP (a|b)=\!^\circ\left(\frac{\tr\lr(ba)}{\tr\lr b}\right)\]
\[P(a)=\!^\circ\left(\frac{\tr\lr a}{\nu}\right).\]

$\lP$ and $\rP$ are called {\it Bayesian conditional probability over\/}
 $Op$. $P$ is
called {\it Bayesian probability over\/} $Op$. 

We see that if $P(b)\neq 0$, $
\lP(a|b)=P(ab)/P(b), \rP(a|b)=P(ba)/P(b).$ Belinfante would call $\lP$ {\it
predictive probability}, and $\rP$ {\it retrodictive probability\/}; $\lP(a|b)$
represents the probability that a measuring operation $a$ of yes-no type
outputs
 ``yes'' at the instant after the observation that $b$ output ``yes'', and
$\rP(a|b)$ represents the probability that $a$ output ``yes''at the instant 
before the observation that $b$ outputs ``yes''. In the
following, ``$\leftarrow$'' reeds ``predictive'' and ``$\rightarrow$'' reeds
 ``retrodictive''.

{\bf Proposition 4.8} $\lP,\rP$ and $P$ have the following properties for
any $a,b,c\in Op$.
\begin{list}{}{}
\item (i) $0\le\lP(a|b),\rP(a|b),P(a)\le 1$ if $b\neq 0$
\item (ii) $\lP(ab|c)=\lP(b|c)\lP(a|bc)$ if $c,bc\neq 0$.
\item (iii) $\rP(ab|c)=\rP(a|c)\rP(b|ca)$ if $c,ca\neq 0$
\item (iv) $\lP(a+b|c)=\lP(a|c)+\lP(b|c)$ if $c\neq 0$
\item (v) $\rP(a+b|c)=\rP(a|c)=\rP(b|c)$ if $c\neq 0$
\item (vi) $P(a+b)=P(a)+P(b).$
\end{list}

{\bf Proof} (i) follows from Proposition 4.6. (ii)--(vi) are evident from
the definition.$\hspace{5mm}\Box$

{\bf Theorem 4.9} (Bayes-type Theorem) Suppose $K\in\N,a_1,...,a_K,b\in
Op$ and $\sum_{k=1}^Ka_k=1$. If $P(b)\neq 0$ and $a_1,...,a_K\neq 0$, then
\begin{list}{}{}
\item (i)$\quad\rP(a_j|b)=\displaystyle \frac{\lP(b|a_j)P(a_j)}{\displaystyle
 \sum_{k=1}^K\lP(b|a_k)P(a_k)}$
\item (ii) $\lP(a_j|b)=\displaystyle \frac{\rP(b|a_j)P(a_j)}{\displaystyle
\sum_{k=1}^K\rP(b|a_k)P(a_k)}$
\end{list}

{\bf Proof} By Proposition 4.8.$\hspace{5mm}\Box$

If $a\neq 0,\lP(\cdot|a),\rP(\cdot|a)$ and $P(\cdot)$ are finitely additive.
However, they do not have $\sigma$-additivity. We also see that Theorem 4.9
does not hold if we let $K$ be an infinite number 
(i.e.,$K\in\us\N\setminus\N$).
$\sigma$-additivity will be argued later.

{\bf Proposition 4.10} 
\begin{list}{}{}
\item (i) $\rP(1|a)+\lP(1|a)=1$ if $a\neq 0$
\item (ii) $\rP(0|a)=\lP(0|a)=0$ if $a\neq 0$
\end{list}
Let $u\in Op$ be a unitary operation and suppose $b\neq 0.$
\begin{list}{}{}
\item (iii) $\lP(uau^{-1}|uau^{-1})=\lP(au^{-1}|ub)=\lP(ua|b)=\lP(a|bu)=
\lP(a|b)$
\item (iv) $\rP(uau^{-1}|ubu^{-1})=\rP(ua|bu^{-1})=\rP(au|b)=\rP(a|ub)
=\rP(a|b).$
\end{list}

 From (i),(iii) and (iv), we see that $\lP(a|u)=\rP(a|u)=P(a)$ and $\lP(u|d)=
\rP(u|b)=P(u)=1$. (iii) and (iv) are called the unitary invariance of $\lP$ 
and $\rP$.

{\bf Theorem 4.11} Suppose $a,b\in Op$ and $b\neq 0$
\begin{list}{}{}
\item (i) $\lP(a|b)=\rP(a\ud|b\ud)$
\item (ii) $\rP(a|b)=\lP(a\ud|b\ud)$
\item (iii) $P(a)=P(a\ud)$
\end{list}

{\bf Proof} (i)--(ii) follow from the fact that if $a\in Op$ then
$\tr\lr a=\tr\lr a\ud.\hspace{5mm}\Box$

\begin{sloppypar}
{\bf Corollary} Suppose $a_m,b_n\in Op$ and $a_m\ud =a_m,b_n\ud =b_n\,
(m=1,...,M,n=1,...,N;M,N\in\us\N)$.
\begin{list}{}{}
\item (i) $\lP(a_1\cdots a_M|b_1\cdots b_N)=\rP(a_M\cdots a_1|b_N\cdots b_1)$
\item (ii) $P(a_1\cdots a_M)=P(a_M\cdots a_1)$.
\end{list}
where if $M$ (resp.$N$) is infinite, $a_1\cdots a_M$ (resp. $b_1\cdots b_N$) 
denotes hyperfinite product of operations.
\end{sloppypar}

Operation $a_M\cdots a_1$ is interpreted as a time series of physical
operations $a_1\rightarrow a_2\rightarrow \cdots \rightarrow a_M$. Thus, if
$a_m\ud=a_m\,(m=1,...,M)$ (e.g., $a_1,...,a_M$ are projecting operations
), we may interpret $a_1\cdots a_M$ as the ``time reversal'' of $a_M\cdots 
a_1$ from this corollary. Generally, $a\ud$ is interpreted as a ``time
reversal'' of $a$ from Theorem 4.11, and hence the theorem shows that we can
describe the time reversal symmetry including measuring processes in the
framework of hyperfinite operations. Thus, this framework is expected 
to complete
Belinfante's program, that is, ``time-symmetric quantum theory'' 
including quantum measurement theory.

\section{Hyperfinite Instrument}
\def\O{{\cal O}}
\def\I{{\cal I}}
\def\J{{\cal J}}
\def\lL{\stackrel{\leftarrow}{L}}
\def\rL{\stackrel{\rightarrow}{L}}
\def\loo{\stackrel{\leftarrow}{\omega}}
\def\roo{\stackrel{\rightarrow}{\omega}}

{\bf Definition 5.1} Let $a_{\xi}\in Op$ ($\xi\in\O,\O$ is a hyperfinite
 set). If function $\I:\O\rightarrow Op$ is internal and 
$\sum_{\xi\in\O}\I(\xi)$ is trivial, $\I$ is called a {\it hyperfinite 
instrument\/}, or 
simply an {\it instrument}. Any element of $\O$ is called an 
{\it outcome\/} of $\I$.

Let $\I:\O_1\rightarrow Op$ and $\J:\O_2\rightarrow Op$ be instruments. The
product $\I\J:\O_1\times\O_2\rightarrow Op$ is defined as
\[(\I\J)(\xi,\zeta)=\I(\xi)\J(\zeta).\]

The product of two hyperfinite instruments is also a hyperfinite instrument,
because
\[\tr\lr\left(\sum_{\xi\in\O_1,\zeta\in\O_2}\I(\xi)\J(\zeta)a\right)
=\tr\lr\Bigl(\sum_{\xi\in\O_1}
\I(\xi)\Bigr)\Bigl(\sum_{\zeta\in\O_2}\J(\zeta)\Bigr)a\]
\[=\tr\lr\left(\sum_{\zeta\in\O_2}\J(\zeta)a \right)=\tr\lr a.\hspace{9em}\]

Let $\I:\O\rightarrow Op$ be a hyperfinite instrument and $a\in Op,a\neq 0
$. If $A$ is an internal subset of $\O$, then $\sum_{\alpha\in A}\I(\alpha)$
 exists.
We define $\lP_{\I},\rP_{\I}$ and $P_{\I}$  as follows:
\[\lP_{\I}(A|a)=\lP\bigl(\sum_{\alpha\in A}\I(\alpha)|a\bigr).\]
\[\rP_{\I}(A|a)=\rP\bigl(\sum_{\alpha\in A}\I(\alpha)|a\bigr).\]
\[P_{\I}(A)=\lP(A|1)=\rP_{\I}(A|1)=P\bigl(\sum_{\alpha\in A}\I(\alpha)\bigr).\]

$\P^i(\O)$, the set of all the internal subsets of $\O$, is a finitely 
additive family, and it is shown that $(\O,\P^i(\O),
\lP_{\I}(\cdot|a))$ is a completely additive probability space. By Hopf's
extension theorem, there exists $\sigma$-additive probability space 
$(\O,\sigma\P^i(\O),\lP_{\I}(\cdot|a))$ which is the extension of 
$(\O,\P^i(\O),\lP(\cdot|a))$, where $\sigma\P^i(\O)$ is the least $\sigma$-
field of sets greater than $\P^i(\O)$. Let $(\O,L\P^i(\O),L\!\lP_{\I}(\cdot|
a))$ denote the Lebesgue completion of $(\O,\sigma\P^i(\O),\lP_{\I}(\cdot|a))$,
and $\lL(\I|a)$ be the abbreviation of it. $\rL(\I|a)=(\O,L\P^i(\O),
L\!\rP_{\I}(\cdot|a))$ is defined in a similar way. Let $L(\I)=(\O,L\P^i(\O),
LP_{\I}(\cdot))=\rL(\I|1)=\lL(\I|1)$.

$\rL(\I|a),\lL(\I|a)$ and $L(\I)$ are called the {\it Loeb probability space
generated by\/} $\I$ (general theory of Loeb space and its application are seen
in, e.g., Ref.(13)).

{\it Bayesian conditional probability\/} in terms of instruments is defined 
as follows.
Let $\I:\O_1\rightarrow Op$ and $\J:\O_2\rightarrow Op$ be instruments.
If $A\in L\P^i(\O_1),B\in L\P^i(\O_2)$ and $LP_{\J}(B)>0,$
\[\lP_{\I,\J}(A|B)=LP_{\I,\J}(A\times B)/LP_{\J}(B),\]
\[\rP_{\I,\J}(A|B)=LP_{\J,\I}(B\times A)/LP_{\J}(B),\]
These are well-defined because it is known that if $A\in L\P^i(\O_1)
$ and $B\in L\P^i(\O_2)$ then $A\times B\in L\P^i(\O_1\times \O_2)$ and
$B\times A\in L\P^i(\O_2\times\O_1).$

Note that while $\lP_{\I}(A|a)$ and $\rP_{\I}(A|a)$ are defined if 
$a\neq 0$ (even if the (unconditional) Bayesian probability $P(a)$ 
equals $0$), $\lP_{\I,\J}(A|B)$ and $\rP_{\I,\J}(A|B)$ are defined only 
if $LP_{\I}(B)>0$.

\section{Bayesian State}
\def\A{{\cal A}}

In Bayesian statistics, one of the most foundational concept is that of 
a priori distribution, which is understood as the probability distribution
in which we have the least information or knowledge about a set of objects,
that is, a sample space. In the following, we argue the concept of quantum
Bayesian a priori and a posteriori states in terms of operations and 
instruments. The ideas are as follows. At an actual measurement process,
or generally, at an experimental operation process,
we often do not know the initial (resp. final) state of objects. In such a
case, we do not
predict (resp. retrodict) the outcome from the initial (resp. final) state,
 but, conversely, guess the initial (resp. final) state from the outcome. 
The Bayesian a priori (resp. a posteriori) state of the 
process is the initial (resp. final) state that we guess in such a way. 

The definition
of the concept uses the following usual definition of states over $C^*$-
algebras$^{(2)}$.

{\bf Definition 6.1} Let $\A$ be a $C^*$-algebra. We define the norm of a
functional $f$ over $\A$ by 
\[\parallel f\parallel =\sup\{|f(A)|:\parallel A\parallel =1\}\]
A linear functional $\omega$ over $\A$ is defined to be positive if
\[\omega(A^*A)\ge 0\]
for all $A\in\A$. A positive linear functional $\omega$ over $\A$ with 
$\parallel\omega\parallel=1$ is called a {\it state\/}. 

It is shown that if
$\A$ contains a unit $1_{\A}$, then $\parallel\omega\parallel=1$ iff 
$\omega(1_{\A})=1$ (see Ref.(14),p.49).

\def\hM{\hat{\M}_{\infty}}
Let $\A=\hM=\fin_{\infty}(\M)/\mu_{\infty}(0). $ $(\hM,^{\circ}\parallel\cdot
\parallel_{\infty})$ is a $C^*$-algebra as we have already seen at section 2.

{\bf Definition 6.2} Let $a\in Op$ and $a\neq 0$. States $\loo_a$ and $\roo_a$
 over $\hM$ are defined as follows:
\[\roo_a(\hat{A})=\!^\circ\left(\frac{\tr a(A)}{\tr\lr a}\right)\]
\[\loo_a(\hat{A})=\!^\circ\overline{\left(\frac{\tr a\ud(A^*)}{\tr\lr a}
\right)}\]
where $\hat A=A+\mu_{\infty}(0).$ $\roo_a$ and $\loo_a$ are called 
{\it Bayesian
 a priori state\/} and {\it Bayesian a posteriori state of \/} $a$, 
respectively. That $\roo_a$ and $\loo_a$ are well-defined is shown as follows.

\begin{sloppypar}
Suppose $A,B\in\fin_{\infty}(\M)$ and $A-B\in\mu_{\infty}$. Notice that 
$\tr [a(A)]=\overline{\tr[A^*a\ud(I)]}$ and $\overline{\tr [a\ud(A^*)]}=
\tr[Aa(I)]$. Hence,
\[\frac{\tr [a(A)]}{\tr\lr a}-\frac{\tr [a(B)]}{\tr\lr a}=\frac{\tr [a(A-B)]}
{\tr[a\ud(I)]}=\tr\left(\frac{(A-B)^*a\ud(I)}{\tr[a\ud(I)]}\right).\] 
$a\ud(I)/\tr[a\ud(I)]$ is positive and its trace is one, that is, $a\ud(I)/
\tr[a\ud(I)]$ is a density matrix. Let $|\xi_1\rangle,...,|\xi_{\nu}\rangle$ 
be the normalized eigenvectors of $a\ud(I)/\tr[a\ud(I)]$, and 
$\lambda_1,...,\lambda_{\nu}$
 be the eigenvalues which correspond to them, respectively. 
$|\xi_1\rangle,...,|\xi_{\nu}\rangle$ is a complete orthonormal system of  
$\us\C^*$.
 Because $(A-B)^*\in\mu_{\infty}(0)$ ,$\lambda_k\ge 0$ and $\sum_{k=1}^{\nu}
\lambda_k=1,$
\[|\tr(A-B)^*a\ud(I)/\tr[a\ud(I)]| \]
\[=|\displaystyle\sum_{k=1}^{\nu}\langle\xi_k|(A-B)^*\frac{a\ud(I)}
{\tr[a\ud(I)]}|\xi_k\rangle|\]
\[\le\displaystyle\sum_{k=1}^{\nu}\lambda_k|\langle\xi_k|
(A-B)^*|\xi_k\rangle|\approx 0.\]
Therefore, $\roo_a(\hat A)=\roo_a(\hat B).$  $\loo_a(\hat A)=\loo_a(\hat B)$
is shown in a similar fashion.
\end{sloppypar}

{\bf Definition 6.3} Let $\I:\O\rightarrow Op$ be a hyperfinite instrument
and $S$ an internal subset of $\O$. States $\roo_{\I,S}$ and $\loo_{\I,S}$
over $\hM$ are defined as follows:
\[\roo_{\I,S}(\hat A)=\!{}^{\circ}\left(\frac{\sum_{\alpha\in S}\tr [\I
(\alpha)(A)]}
{\sum_{\alpha\in S}\tr\lr\I(\alpha)}\right)\]
\[\loo_{\I,S}(\hat A)=\!{}^{\circ}\overline{\left(\frac{\sum_{\alpha\in S}
\tr [\I(\alpha)\ud(A^*)]}{\sum_{\alpha\in S}\tr\lr\I(\alpha)}\right)}\]
$\roo_{\I,S}(\,\loo_{\I,S}\,)$ is called {\it Bayesian a priori (a posteriori)
 state of\/} $\I$ when its outcome is in $S$.

Note that if $a=\sum_{\alpha\in S}\I(\alpha),$ then $ \roo_{\I,S}(\hat A)
=\roo_a(\hat A), \loo_{\I,S}(\hat A)=\loo_a(\hat A)$, and hence $\roo_{\I,S}$
and $\loo_{\I,S}$ are well-defined.

We have following relations concerning $\lP,\rP,\loo,\roo.$ Let $a\in Op$
satisfy $a(A)=\sum_{k=1}^{\kappa}M_kAM_k^*$ for all $A\in\M$ where $\kappa\in
\us\N,$ $ M_1,...,M_{\kappa}\in\M$ and $\sum_{k=1}^{\kappa}M_k^*M_k\le I,$ $
\sum_{k=1}^{\kappa}M_kM_k^*\le I.$ Let $b\in Op$ and $b\neq 0.$ We have
\[\lP(a|b)=\loo_b(\hat M) \]
\[\rP(a|b)=\roo_b(\hat M')\]
where $M=\sum_{k=1}^{\kappa}M_k^*M_k,$ $ M'=\sum_{k=1}^{\kappa}M_kM_k^*.$

\section{Remarks and Problems}

1. This paper presents no content but a framework of time-symmetric quantum
physics. To describe the physical content in the framework, we first need
representations of canonical commutation relation on hyperfinite-dimensional
linear spaces. Ojima and Ozawa$^{(15)}$ started the research in this 
direction. 

2. Some mathematical problems are left. First, the condition of the definition
of $\lP_{\I,\J}$ and $\rP_{\I,\J}$, that is, $LP_{\J}(B)>0$ should be 
weakened. $\lP_{\I,\J}(A|B)$ and $\rP_{\I,\J}(A|B)$ seem not to be 
always meaningless
even if $LP_{\J}(B)=0$. Secondly, Bayesian state of an instrument is
defined only when $S$ is internal in Definition 6.3. Can we extend the 
definition so that it contains the cases that $S$ is external?  
Thirdly, can we define Bayesian a priori and 
a posteriori states as normal states on von Neumann algebra $\hat\M$? 
\vspace{10mm}

{\bf REFERENCES}
\begin{description}
\item 1. F.J.Belinfante, {\it Measurement and Time Reversal in Objective
Quantum Theory}, Oxford:Pergamon Press(1975).
\item 2. R.Haag and D.Kastler, ``An Algebraic Approach to Quantum Field
Theory,'' {\it J.Math.Phys.}5(7),848--861(1964).
\item 3. R.Haag, {\it Local Quantum Physics}, New York:Springer(1992).
\item 4. J.von Neumann, {\it Mathematical Foundations of Quantum Mechanics},
Princeton:Princeton University Press(1955).
\item 5. M.Ozawa, ``Cat Paradox for $C^*$-Dynamical Systems,'' 
{\it Prog.Theo.Phys.}\\88(6),1051--1064.(1992)
\item 6. E.B.Davies,and J.T.Lewis, ``An Operational Approach to Quantum
Probability,'' {\it Commun.Math.Phys.}17,239--260(1970).
\item 7. E.B.Davies, {\it Quantum Theory of Open Systems\/}, London:Academic
Press\\(1976)
\item 8. A.Robinson. {\it Non Standard Analysis},Amsterdam:North-Holland(1966).
\item 9. A.E.Hurd, and P.A.Loeb, {\it An Introduction to Nonstandard
Real Analysis}, Orlando:Academic Press(1985).
\item 10.T.Hinokuma and M.Ozawa, ``Conversion from Nonstandard Matrix Algebras
to Standard Factors of Type II$_1$,'' {\it Illinois Journal of Mathematics}
37(1),1--13(1993).
\item 11. M.D.Srinivas, ``Foundations of a Quantum Probability Theory,''
{\it J.Math.\\Phys.}16(8)(1975).
\item 12. M.Ozawa, ``Realization of Measurement and the Standard Quantum
Limit,'' in {\it Squeezed and Nonclassical Light\/}, ed. by P.Tombesi and
E.R.Pike. Plenum(1989).
\item 13. S.Albeverio, J.Fenstad, R.H{\o}egh-Krohn, and T.Lindstr{\o}m,
{\it Nonstandard Methods in Stochastic Analysis and Mathematical Physics},
Orlando:Academic Press(1986).
\item 14. O.Bratteli and D.W.Robinson, {\it Operator Algebras and Quantum
Statistical Mechanics I},''(second edition), New York:Springer(1987).
\item 15. I.Ojima and M.Ozawa, ``Unitary Representations of the Hyperfinite
Heisenberg Group and the Logical Extension Methods in Physics'', {\it Open
Systems and Information Dynamics\/}2(1),107-128(1993).
 
\end{description}

\end{document}